\documentclass[final,5p,times,twocolumn]{elsarticle}

\usepackage{epstopdf}
\usepackage{epsfig}
\usepackage{graphicx}
\usepackage{subfigure}
\usepackage{amsthm}
\usepackage{lineno}
\usepackage[tableposition = top]{caption} 
\usepackage{booktabs}                     
\usepackage[per-mode = symbol]{siunitx}   
\usepackage{hyperref}

\biboptions{sort&compress}

\journal{Journal of Molecular liquids}

\begin{document}

\begin{frontmatter}

\title{Suspensions of supracolloidal magnetic polymers: self-assembly properties from computer simulations}

 \author[label1]{Novak E.V.}
 \author[label1]{Pyanzina E.S.}
 \author[label2]{Rozhkov D.A.}
 \author[label2]{Ronti M.}
 \author[label3]{Cerd{\`a} J. J.}
 \author[label4]{Sintes T.}
 \author[label1,label2]{S\'anchez P.A.}
 \author[label1,label2]{Kantorovich S.S.}
 \address[label1]{Ural Federal University, Lenin Av. 51, Ekaterinburg, 620000, Russia}
 \address[label2]{University of Vienna, Sensengasse 8, 1090 Vienna, Austria}
  \address[label3]{Departament de F\'isica, Universitat de les Illes Balears, E-07122, Palma de Mallorca, Spain}
  \address[label4]{Instituto de F\'isica Interdisciplinar y Sistemas complejos, IFISC (UIB-CSIC), E-07122, Palma de Mallorca, Spain}

\begin{abstract}
We study self-assembly in suspensions of supracolloidal polymer-like structures made of crosslinked magnetic particles. Inspired by self-assembly motifs observed for dipolar hard spheres, we focus on four different topologies of the polymer-like structures: linear chains, rings, Y-shaped and X-shaped polymers. We show how the presence of the crosslinkers, the number of beads in the polymer and the magnetic interparticle interaction affect the structure of the suspension. It turns out that for the same set of parameters, the rings are the least active in assembling larger structures, whereas the system of Y- and especially X-like magnetic polymers tend to form very large loose aggregates.   
\end{abstract}


\end{frontmatter}

\section{Introduction}

Nowadays, the creation of smart materials  relies on a multiscale design, from the nanoscale to macroscopic properties. The internal structure at the nano- and micro-levels determines the texture, elasticity, viscosity, taste and other macroscopic properties of soft materials. There are several techniques to change the properties of soft materials: by varying the pH balance, temperature, turning on and off the external fields. The essential condition to use a magnetic field as a control  parameter is the presence of magnetically sensitive components in a soft material. There are several ways to incorporate such components into liquids and gels. The common thing for all the techniques is the size of the magnetic building blocks – magnetic colloids in the range from a couple of nanometers to several microns. Magnetic colloids in liquid or elastic carriers, directed by applied magnetic fields, or under the action of intrinsic magnetic forces, exhibit hierarchical self-assembling and various structural-phase transitions, which, in turn, can lead to macroscopic changes of all soft material. The list of possible structures and phases is very large and is determined by the size, concentration, type and material of magnetic inclusions. 

The oldest and, probably, the most understood example of magnetic soft matter is a ferrofluid \cite{resler64a}, {\it i.e.} a system of surface-stabilised single-domain magnetic nanoparticles suspended in a magnetopassive carrier. Nanoparticles in this systems are known to self-assemble \cite{klokkenburg06a,camp00a,weis93a,1970-degennes,pshenichnikov96b} and through clustering affect strongly viscous \cite{odenbach02b,odenbach02a}, optical \cite{scholten80a,taketomi83a,hasmoney00a}, magnetic \cite{2013-kantorovich-prl,klokkenburg08a} and diffusion properties \cite{dobroserdova17a,erne03a,1995-bacri}. Even though the self-assembly of magnetic nanoparticles seems to be a promising tool to control the response of a ferrofluid, such structural transformations are very sensitive to noise created by temperature fluctuations \cite{morales09a}, particle polydispersity \cite{munoz-menedez15a,2004-ivanov} or particle asphericity \cite{2014-tierno,donaldson17a}.

One of the avenues to avoid such a sensitivity of self-assembly is to predefine the structural motifs: to crosslink the magnetic particles in so-called magnetic filaments \cite{2005-dreyfus, 2008-erglis-jpcm, 2008-corr,2017-novak-jmmmb,2018-kuznetsov-jmmm, 2018-hernandez-rojas-pre} or other polymer-like supracolloidal structures \cite{rozhkov17a}. In this case, cluster sizes and shapes cannot be altered by temperature and such clusters will remain connected even under conditions for which self-assembly in a ``regular ferrofluid'' would have not taken place. However, the question arises: ``will, and, in case, how, supracolloidal structures self-assemble?''

In the present computer simulation study, we investigate suspensions of supracolloidal magnetic polymer-like structures (SMP) of linear (LSMP), ring (RSMP), Y- (YSMP) and X-shapes (XSMP), since these structures are predominant at low temperatures in systems of dipolar fluids \cite{2013-kantorovich-prl, 2000-safran, 2000-ilg1}. We vary the length of SMPs, that is the number of magnetic particles forming them; the concentration of SMPs in the suspension; and the strength of magnetic interparticle interactions. Additionally, we perform the analysis of a ferrofluid with non-crosslinked magnetic particles under the same set of conditions. In this way, we do not only elucidate the influence of crosslinkers on the hierarchical self-assembly, but can also envision the topology-driven structural transitions. We found that while RSMPs are inert and do not self-assemble, LSMPs under the same conditions can exhibit cluster formation, albeit not as strong as that found for  YSMPs and XSMPs.

The structure of the manuscript is the following: firstly, in section \ref{sec-model}, we discuss computational methods used to study SMPs; next, we present results on cluster-size distributions for various SMPs (section \ref{sec-csd}), analyse how the position of the magnetic bead in a SMP influences its ability to form a connection (section \ref{sec-maps}), and describe the topology of SMP clusters, looking at the types inter-SMP bonds (section \ref{sec-bond}); finally, we summarise our work in section \ref{sec-conc}.

\section{Model and simulation details}\label{sec-model}

In this work we employ computer simulations with a bead-spring model in order to study the self-assembly of dispersions of SMPs with different parameters, and compare their properties with the ones corresponding to analogous dispersions of free dipolar particles (\textit{i.e.}, pure model ferrofluids). In order to model a pure ferrofluid we consider $N$ ferromagnetic spherical particles with diameter $\sigma=1$ and mass $m=1$. Each particle has a magnetic moment, $\vec \mu$, in its centre. Interactions between particles in such a system are described by a combination of two potentials. The first one is the dipole-dipole potential, that models the long-range magnetic interaction between any pair $i$, $j$ of magnetic particles:
\begin{equation}
U_{dd}(\vec r_{ij}; \vec \mu_1, \vec \mu_2)=\frac{\vec{\mu}_{i}\cdot\vec{\mu}_{j}}{r^{3}}-\frac{3\left[\vec{\mu}_{i}\cdot\vec{r}_{ij}\right]\left[\vec{\mu}_{j}\cdot\vec{r}_{ij}\right]}{r^{5}},
\label{eq:dipdip}
\end{equation}
where $\vec \mu_i$ and $\vec \mu_j$ are their respective dipole moments, $\vec r_{ij} = \vec r_i - \vec r_j$ is the displacement vector connecting their centres and $r=\left \| \vec r _{ij}\right \|$. 
The second one, is the soft core interaction between these particles, described by the Weeks-Chandler-Andersen pair potential \cite{1971-weeks}:
\begin{equation}
U_{\mathrm{{WCA}}}(r)=\left\{ \begin{array}{ll}
U_{\mathrm{{LJ}}}(r)-U_{\mathrm{{LJ}}}(r_{cut}), & r<r_{\mathrm{{cut}}}\\
0, & r\geq r_{\mathrm{{cut}}}
\end{array}\right. ,
\label{eq:WCA}
\end{equation}
where $U_{\mathrm{{LJ}}}(r)$ is the conventional Lennard-Jones potential, $U_{\mathrm{{LJ}}}(r)=4 \left ( r^{-12}-r^{-6} \right )$, that in (\ref{eq:WCA}) has been made purely repulsive by truncating and shifting it at the position of its minimum, $r_{\mathrm{{cut}}}=2^{1/6}$.

\begin{figure}[!h]
\centering
\includegraphics[width=5cm]{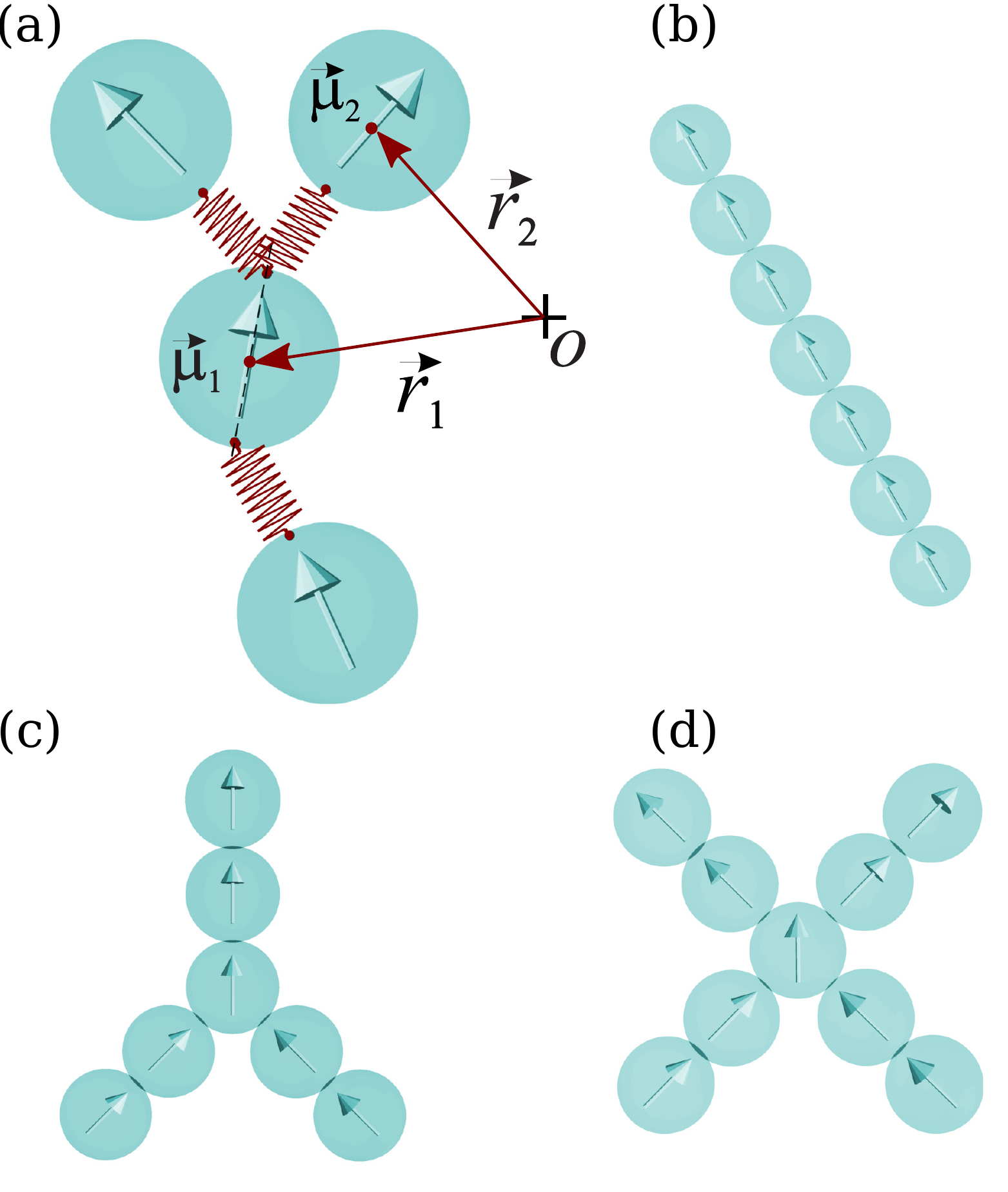}
\caption{Scheme of the term of the bonding potential that couples the orientation of the dipoles and the chain backbone (a) and the main studied SMP topologies: (b) LSMP; (c) YSMP; (d) XSMP.}
\label{fig:model}
\end{figure}

For simulating SMPs, additionally to steric and magnetic interactions, we take into account the permanent bonds between magnetic particles established by the polymer crosslinkers, using a spring-like bonding potential consisting of two terms. The first one is a simple harmonic spring whose ends are attached to the surface of the bonded particles. The spring attachment points are located at the projection points of the head and the tail of the central dipole moment. Figure \ref{fig:model}(a) shows a scheme of this bonding term, that effectively couples the orienation of the dipoles and the chain backbone. The second part of the potential corresponds to a FENE interaction connecting the centres of the linked particles, that limits the maximum extension of the bond. Therefore, the net potential is defined as:
 \begin{equation}
     U_{S}(\vec r_{ij}) = \frac{K}{2} \left [ \left( \vec r_{ij} - \frac{1}{2}(\hat{\mu}_i+\hat{\mu}_j) \right )^2 -\frac{r^2_0}{2}\ln\left[1-\left(\frac {\vec r_{ij}}{r_0} \right)^2 \right] \right ],
     \label{eq:bondmodel}
  \end{equation}
where $K$ is the energy scale of the interaction, $\hat{\mu}_i=\vec \mu_i / \left \| \vec \mu_i\right \|$ and $\hat{\mu}_j=\vec \mu_j / \left \| \vec \mu_j\right \|$ are the unitary vectors parallel to each associated dipole moment and $r_0$ is the maximum allowed extension for the bond. We take the same parameters for the bonding potential used in our previous studies\cite{2017-novak-jmmmb, rozhkov17a}: $K=30$ and $r_0=1.5$ in our reduced units. With this potential we define the different SMP topologies we study, that are also schematised in Figure \ref{fig:model}.

We performed molecular dynamics simulations in the canonical ensemble at reduced temperature $T=1$, using a Langevin thermostat in order to approximate implicitly the effects of the thermal fluctuations of the carrier fluid. In order to mimic a pseudo-infinite system we employed periodic boundary conditions. The long range magnetic interactions were calculated using the dipolar-P$^3$M algorithm \cite{2008-cerda-jcp}. We sampled systems of up to 3200 particles, arranged in up to 160 SMPs. Starting from a random distribution of SMPs or free particles, we first equilibrated the systems for $5 \cdot 10^5$ integration steps and then performed 350 measurements during a production cycle of $4\cdot 10^6$ integration steps, using a time step of $\delta t = 0.005$. We employed the simulation package {ESPResSo} 3.3.1 \cite{2013-arnold}.


\section{Results and Discussions}\label{sec-res}
In systems of dipolar particles like those we study here, the self-assembly of the basic dispersed units (single particles in ferrofluids, single crosslinked motifs in dispersions of SMPs) is driven by the dipole-dipole interactions between particles. According to the crosslinking scheme assumed for the SMPs, the permanent bonds tend to keep a strongly attractive dipole-dipole interaction between crosslinked neighbours along the chains. Additionally, in an analogous way to the aggregation of free dipolar particles in ferrofluids, particles belonging to SMPs can form non permanent connections with other particles, from the same or different SMPs, as long as their centre-to-centre distance is small and the relative orientation of their dipoles is favourable. In difference with permanent bonds, such non permanent connections, similar to those in non-crosslinked systems, can form and break, balancing energetic contribution and thermal fluctuations in thermodynamic equilibrium. Importantly, non permanent connections allow the aggregation of individual SMPs into larger clusters.

Here we study the spontaneous self-assembly of SMPs with linear, Y- and X- topologies by analysing the properties of the clusters they form in absence of external fields. Dispersions of RSMPs were also simulated in our study, showing no significant assembly under any condition, as it was expected \cite{rozhkov17a}. For the rest of topologies, we use three parameters that provide different levels of detail of the structure of their clusters. We start by calculating the probability distributions for the sizes of the clusters. Secondly, we obtain the probability of two given particles to form non permanent connections depending on their positions within the structure of the SMP they belong. Finally, we perform a systematic analysis of the local topology of the connections. In all cases we identify the non permanent connections between the particles by means of a combination of distance and energy criteria: two particles are considered to be connected if their centre-to-centre distance is smaller than $r_{ij} \le 2^{1/6}$ and their dipole-dipole pair energy, given by expression \ref{eq:dipdip}, $U_{dd}(\vec r_{cut}; \vec \mu_1, \vec \mu_2)<0$. For all systems we sampled two particle number densities, $\rho=0.01$ and $\rho=0.05$. We also sampled two values of the squared dipole moment of the particles, $\mu^2=2$ and $\mu^2=5$, that correspond respectively to a weak and a strong dipole-dipole interaction, relative to the strength of the thermal fluctuations. The latter was set by fixing the reduced temperature to $T=1$. As it happens with molecular polymers, the configurational entropy of SMPs depends on the length of their chain-like segments. We analyse this effect by sampling different SMP sizes, that lay within the range $L \in [9,\,21]$. The fact that the topologies we study have different symmetries prevents to set up samples with exactly the same amount of particles per SMP. Therefore, in our discussion we will need to compare systems with slightly different SMP sizes. Finally, in order to compare our results for dispersions of SMPs with analogous systems of simple ferrofluids, we also performed simulations of free dipolar particles with the same sets of dipole moments and number densities.

\subsection{Cluster size distributions} \label{sec-csd}
The most coarse parameter we discuss is the probability distribution, $P(C^*)$, of the relative cluster size in each system, $C^*$. This is computed as the fraction of clusters composed of a given amount of connected SMPs, averaged over all measures and normalised by the total amount of SMPs in the system, $N$. This latter normalisation makes $C^*$ to vary between $1/N \approx 0$ (corresponding to clusters of size 1, \textit{i.e.}, to non aggregated SMPs) and 1 (corresponding to clusters formed by all SMPs in the system, \textit{i.e.}, a fully connected system). In this way a qualitative, system size-independent comparison of the distributions can be easily performed. This is convenient because, as we pointed above, topological constrains do not allow to sample exactly the same sizes for all systems. However, it is important to underline that this comparison can not be simply extended to much larger systems. Specifically, the probability of finding a fully connected system in our samples only indicates what could be the probability of finding clusters with a size of, at least, the same amount of SMPs we sampled here, but not necessarily to the probability of finding a much larger, fully connected system. In other words, in this study we focus on the impact of the topology, SMP size and density on the appearance of significantly large clusters, leaving the analysis of the conditions for the formation of fully connected systems for future studies.
\begin{figure}[!h!t]
\centering
\subfigure[]{\label{fig:clust-hist-linear}\includegraphics[width=7.5cm]{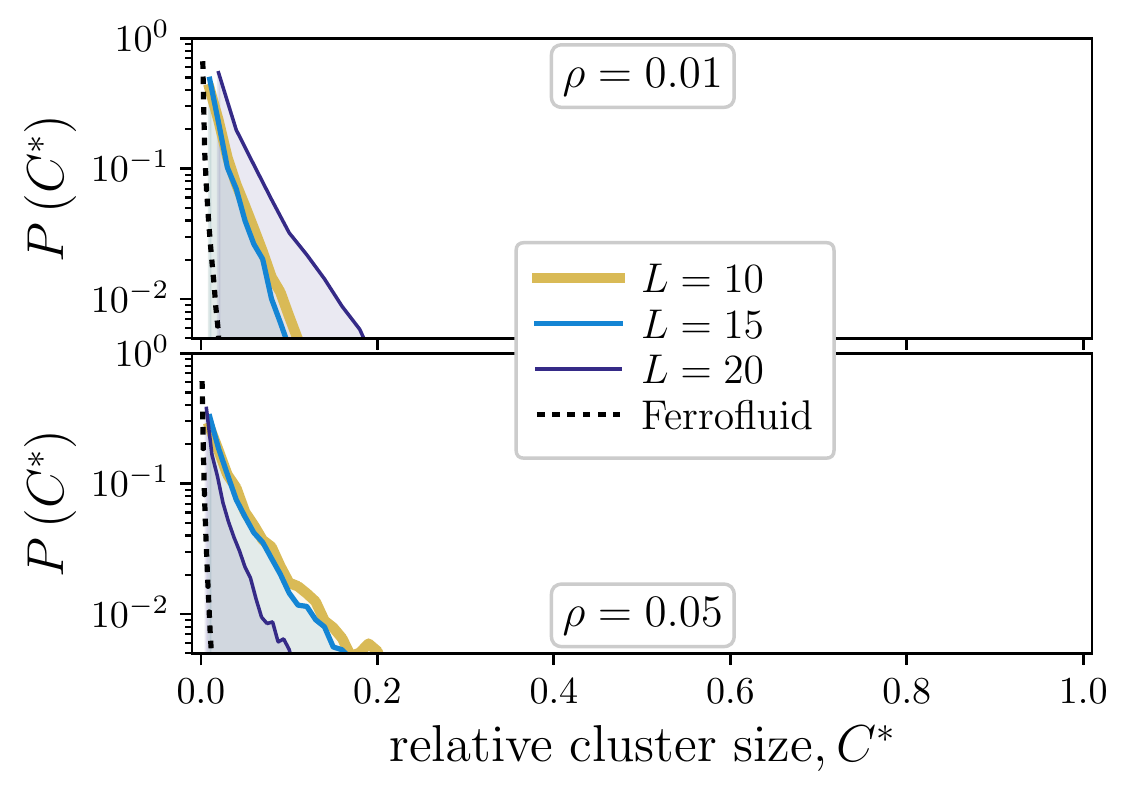}}
\subfigure[]{\label{fig:clust-hist-Y}\includegraphics[width=7.5cm]{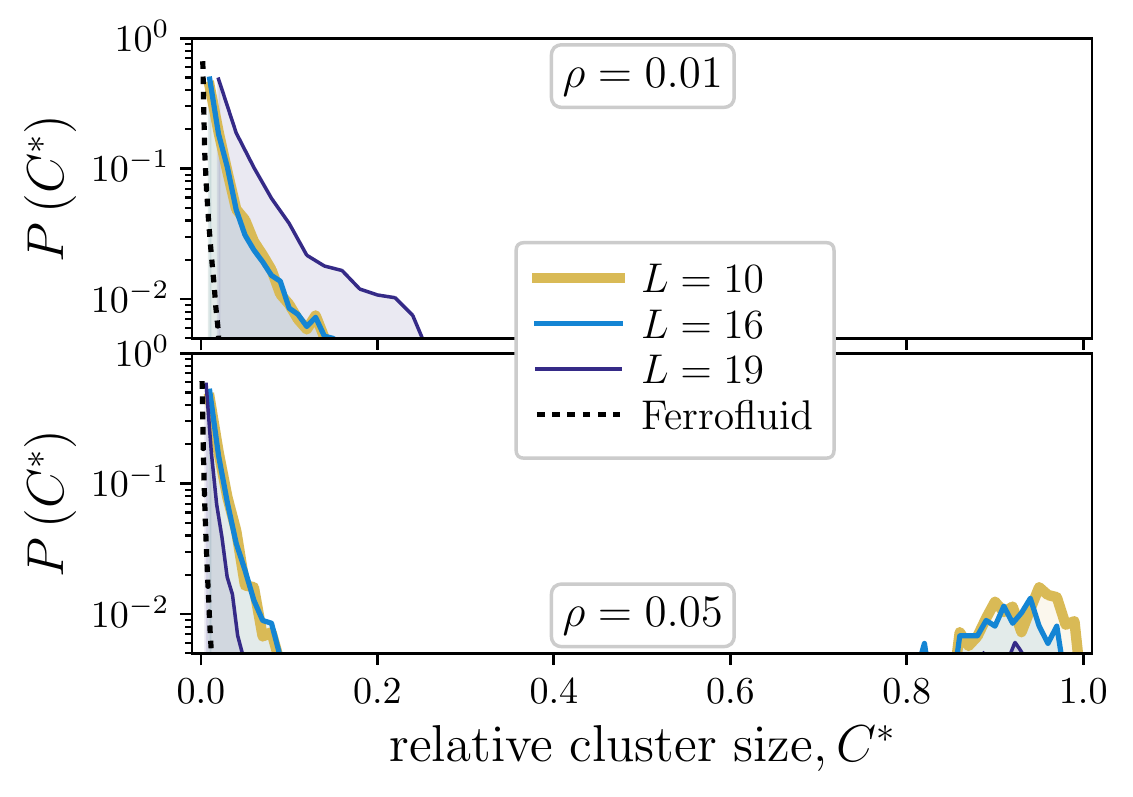}}
\subfigure[]{\label{fig:clust-hist-X}\includegraphics[width=7.5cm]{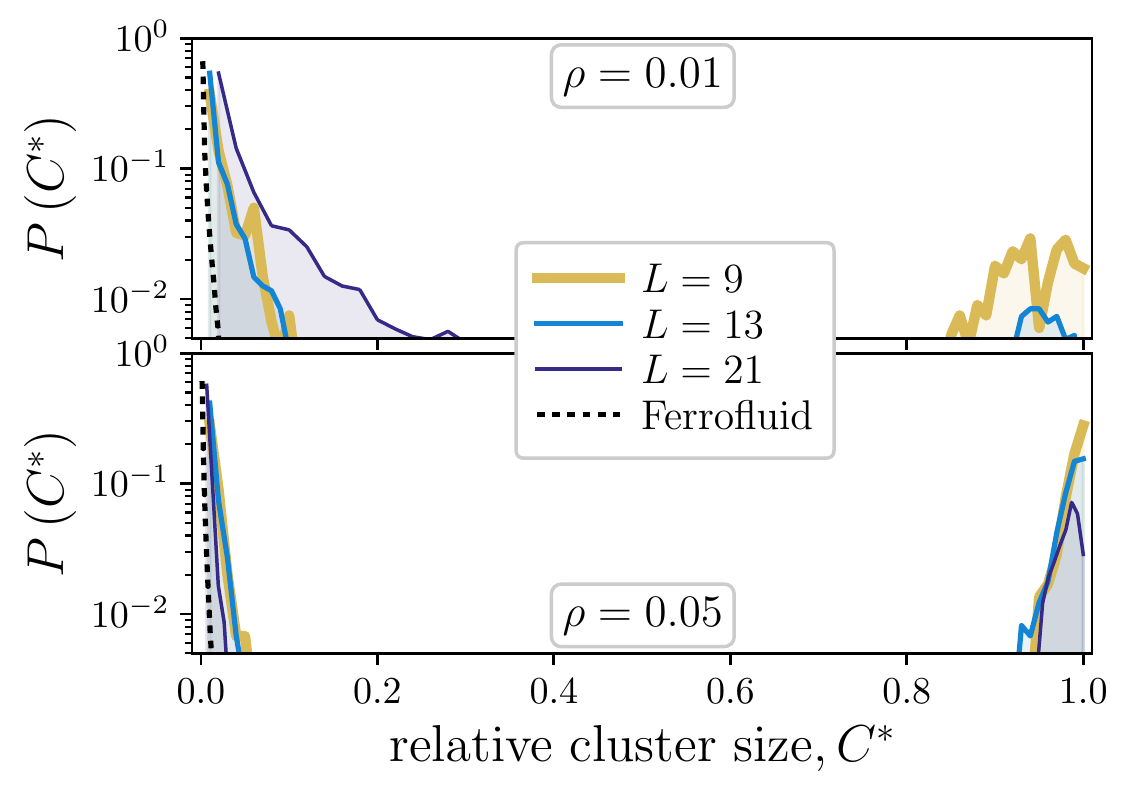}}
\caption{Probability distributions of relative cluster sizes (number of SMPs per cluster divided by the total number of SMPs in the system) with $\mu^2 = 5$. (a) For LSMPs. (b) For YSMPs. (c) For XSMPs. In all cases, relative errors estimated for the leftmost maximum observed in the distributions, $P(C^* \sim 0)$, are not larger than 7\%. For those distributions in which the probability within the region of large cluster sizes, $P(C^* \sim 1)$, is not negligible, relative errors of the corresponding rightmost maximum are below 20\%.}
\label{fig:clust-histograms}
\end{figure}

Figure \ref{fig:clust-histograms} shows a selection of results obtained for this parameter in semilogarithmic scale. Specifically, we show the probability distributions corresponding to LSMPs, YSMPs and XSMPs with strong dipole-dipole interactions, both sampled densities and three selected SMP sizes. In all cases, one can see that there is a large probability of having a significant fraction of isolated SMPs in the system, independently from the topology and the density. This probability decays in a roughly exponential way as one looks for larger cluster sizes. This exponential decay is qualitatively similar to the behaviour corresponding to analogous model ferrofluids, also displayed in these plots as dashed lines. The exponential decay in the distribution of cluster sizes in model ferrofluids with weak and moderate magnetic interactions has been determined in several theoretical works \cite{2000-texeira,2002-wang}. Quantitatively, all SMP systems display a slower decay for small cluster sizes than the equivalent ferrofluid. These distributions also show an interesting interplay between the size of the SMPs and the density of the system in the small cluster size region: whereas for low density the decay corresponding to the largest SMPs is the slowest, at high density it becomes the fastest. Besides this significant dependence on the density and SMP size, that also manifests in the region of large clusters, $P(C^*) \approx 1$, the most striking impact on the distributions is the one coming from the SMP shape. This can be seen by comparing Figures \ref{fig:clust-hist-linear} (corresponding to LSMPs), \ref{fig:clust-hist-Y} (YSMPs) and \ref{fig:clust-hist-X} (XSMPs): while LSMPs show a simple distribution with the discussed almost exponential decay, for YSMPs, at high values of $\rho$, it also appears a significant probability of finding very large clusters, close to the size corresponding to the total amount of SMPs in the system. This tendency to exhibit a bimodal cluster-size  distribution becomes even more pronounced for XSMPs, that display this effect even at low $\rho$. The extreme case corresponds to XSMPs with the smallest size and the highest density, for which the two maxima of the bimodal distribution are almost equivalent: this represents a system composed of very few isolated XSMPs, with the rest of them being aggregated into a single large cluster. These observations support the predictions about SMP dispersions that we made from the indications obtained for SMP pairs \cite{rozhkov17a}: the topology of SMPs and, particularly, the amount of their free ends, increases dramatically the degree of self-assembly of these systems. Finally, it is also worth to mention that all the results obtained for weak dipole-dipole interactions (not shown) also exhibit a simple exponential decay of the cluster size, still slower than the one corresponding to an equivalent model ferrofluid.

The discussion above allows us to extract two main general conclusions. Firstly, SMPs tend to form larger clusters than equivalent ferrofluids. Second, under strong dipole-dipole interaction conditions, as we increase the concentration of magnetic particles or/and enlarge the amount of free ends or `valence' of the SMP topology and/or reduce the SMP size, it is more probable to find very large clusters, with a size that compares to the total amount of SMPs in the system.

\begin{figure}[!t]
\centering
\subfigure[]{\label{fig:maps-Y}\includegraphics[width=7.8cm]{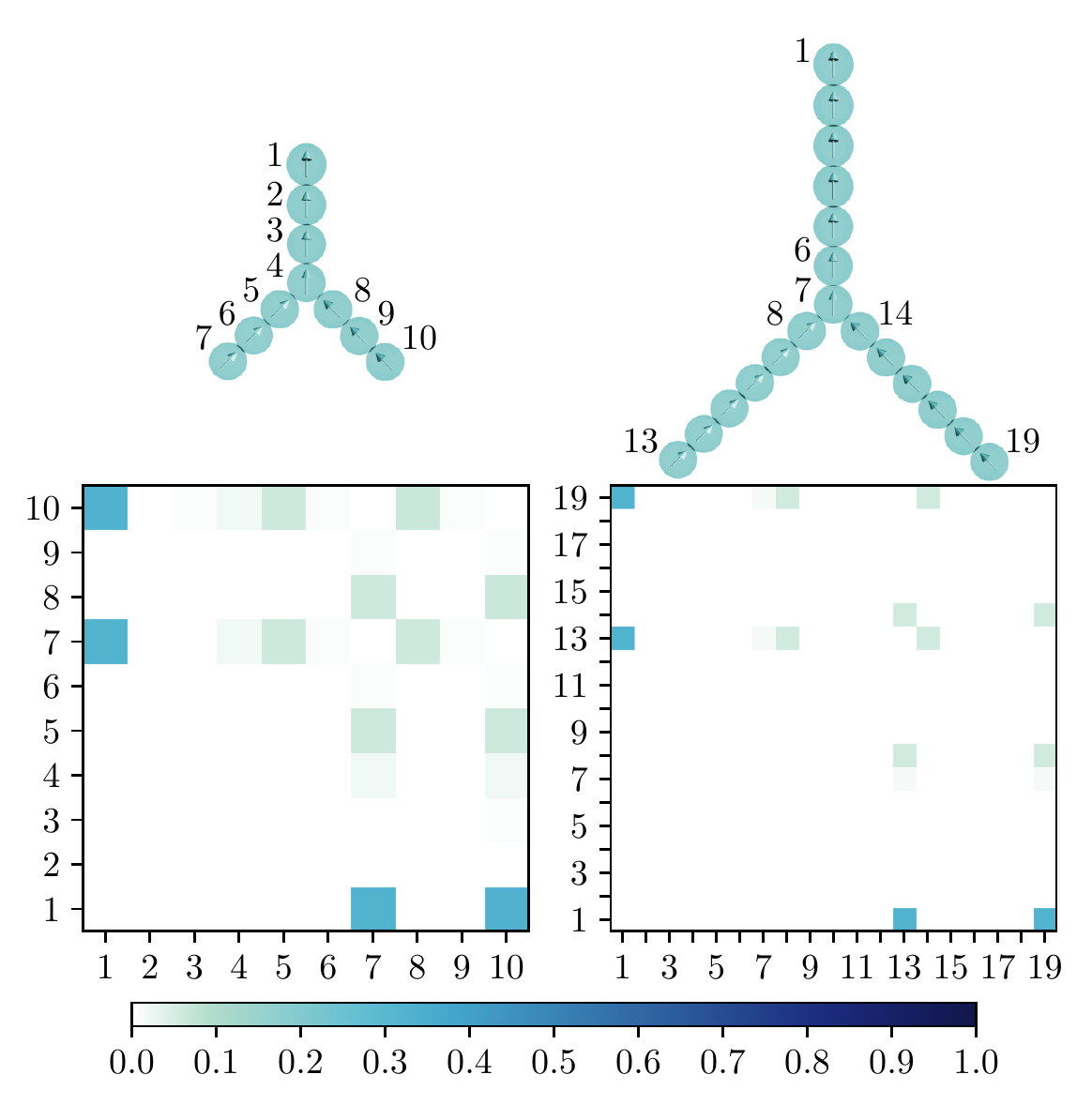}}
\subfigure[]{\label{fig:maps-X}\includegraphics[width=7.8cm]{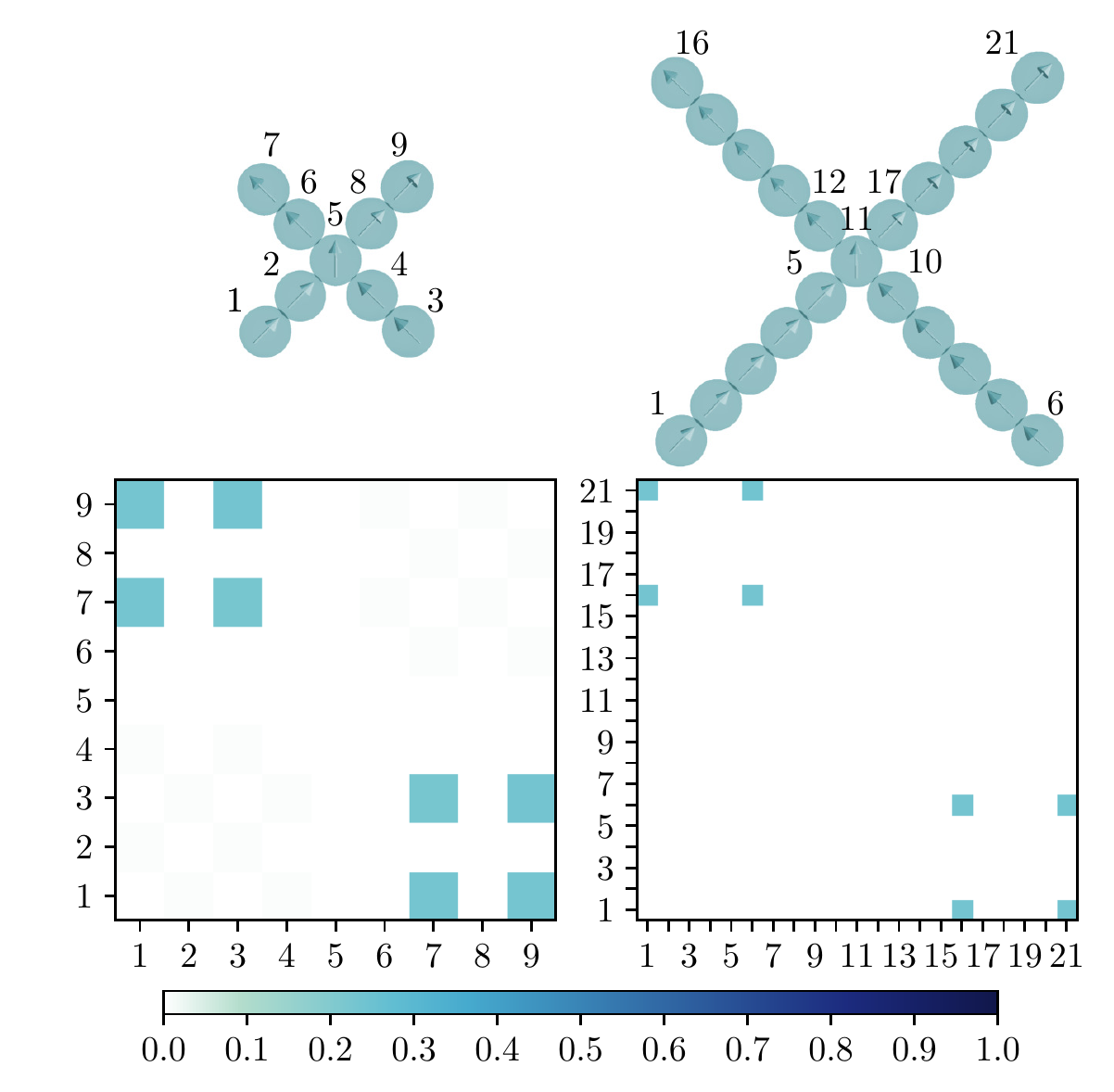}}
\caption{Connectivity probability maps for different SMP dispersions with $\mu^2=5$, $\rho=0.05$ and two selected lengths. Labels of the particles according to their position in the SMP are depicted in the upper row. (a) YSMPs with $L=10$ (left) and $L=19$ (right). (b) XSMPs with $L=9$ (left) and $L=21$ (right).}
\label{fig:connmaps}
\end{figure}
\subsection{Connectivity maps} \label{sec-maps}
The interplay between the strongly directional dipole-dipole interaction and the predefined topology of the SMPs limits the possibilities to establish favourable connections. The overall properties of the aggregates made of SMPs are largely determined by those particles whose positions in SMPs allow to form extra connections. Here, this effect is analysed in the following way. We choose two particles form different SMPs and calculate the probability for them to form a connection, depending on their geometrical location inside these SMPs. By labelling the particles with an index according to their position, we obtain two-dimensional symmetric probability arrays in which each element $P_{ij}$ shows the probability of particles at positions $i$ and $j$ to be connected. These arrays can be represented graphically as connectivity maps using a colour scale for the probability values.

Figure \ref{fig:connmaps} shows a selection of connectivity maps obtained from our simulations. They correspond to dispersions of YSMPs and XSMPs with strong dipole-dipole interaction, the highest value of $\rho$ and the two extreme sampled SMP sizes. Each connectivity map includes a scheme with the labelling of the particles according to their position. In order to ease the discussion, at this point it is convenient to distinguish the main orientation of the particle's dipole at the free ends and around the junction points of each topology: we define a free end to have `a dipole out' when it is not bonded to a neighbour at the point corresponding to the head of its dipole (for example, particle 1 in YSMPs, see upper row of Figure \ref{fig:maps-Y}); analogously, we define a free end to have `a dipole in' when the point corresponding to the tail of its dipole is not bonded; bonded neighbours of junction particles can be also defined as dipoles `in' or `out' depending whether the junction particle is bonded to their corresponding head or tail points, respectively. For example, in the upper left scheme of Figure \ref{fig:maps-Y}, particle 4 is the junction particle, particles 5, 8, 7 and 10 have each a dipole in and particles 1 and 3 have a dipole out.

\begin{figure*}[!h]
\centering
\includegraphics[width=16cm]{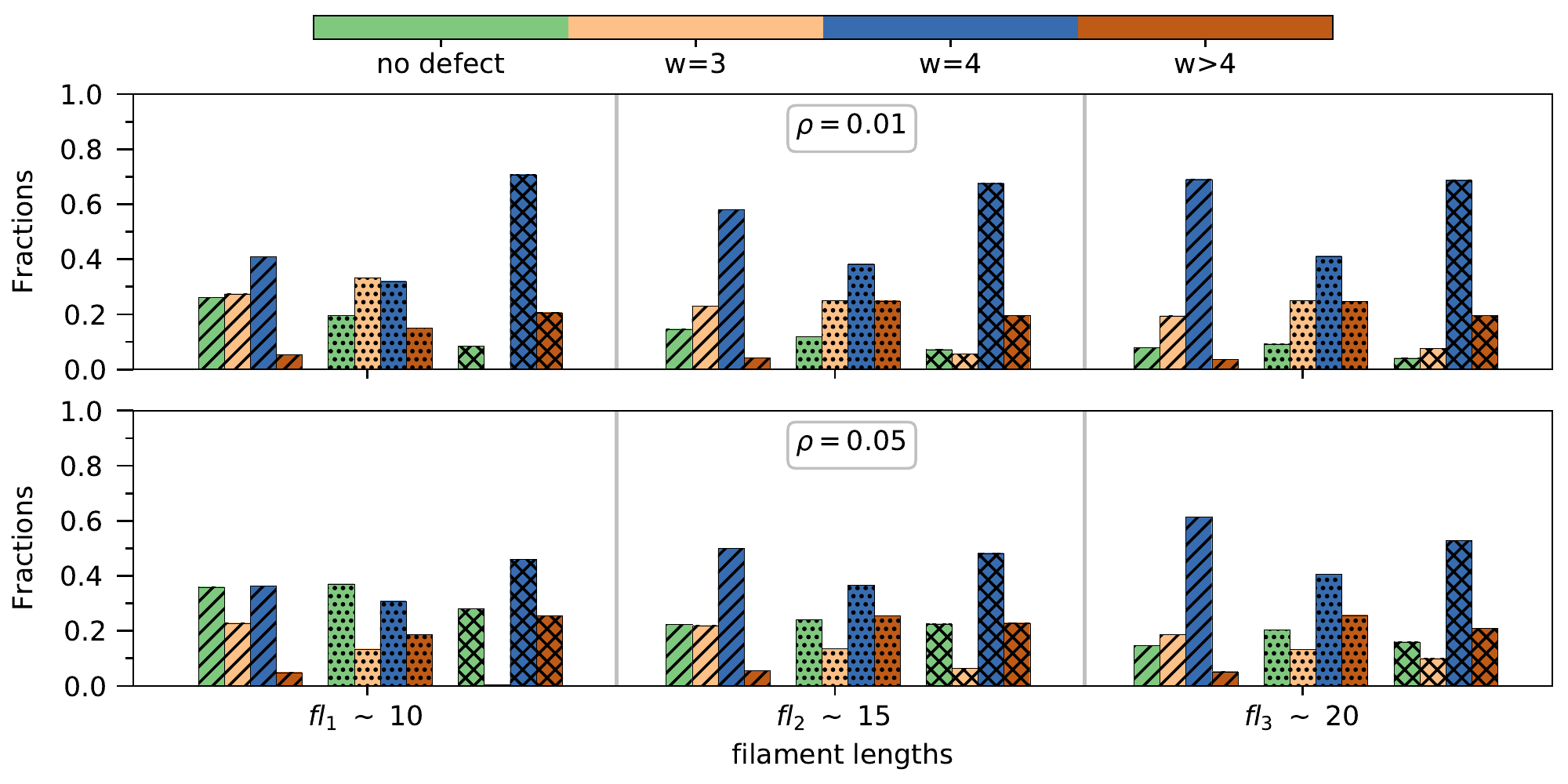}
\caption{\emph{Bar chart $\lambda=2$.} Fractions of particles bonded without the formation of a junction (\emph{green}), with the formation of a $w=3$ defect (\emph{orange}), $w=4$ defect (\emph{blue}) or $w>4$ defect (\emph{red}), for three different values of the filament length. The density value is $\rho=0.01$ for the bar chart on the top, $\rho=0.05$ on the bottom. The hatching of the bars indicates the type of the filaments in the system according to the following identifications: diagonal lines, LSMPs; points, YSMPs; and crosses, XSMPs. Values for simple fluids: for $\rho=0.01$, all the particles form connections without junctions; for $\rho = 0.05$ the fractions are $98.8\%$ particles without junctions, $1\%$ with $w=3$, $0.2\%$ with $w=4$. The maximum relative error estimated for these histograms is lower than 1\%.}
\label{fig:barchartL2}
\end{figure*}

The connectivity maps obtained for YSMPs, shown in Figure \ref{fig:maps-Y}, evidence that the most probable connection in this topology, independently from the SMP size, is established between free ends, as long as one has a dipole in and the other a dipole out. In general, such connections can be very likely because they are compatible with head-to-tail dipole-dipole arrangements. Analogously, connections between free ends with the same dipole orientation are in general energetically very unfavourable. Interestingly, we can see a lower but still significant probability to find connections between free ends with dipole in and particles that have dipoles in respect to the junction. In the case of XSMPs, whose examples of connectivity maps are shown in Figure \ref{fig:maps-X}, the only highly probable connection is the one between free ends with different dipole orientations, whereas junctions and their vicinity play no significant role. This suggests that, in any system with strong dipole-dipole interactions, connections between compatible free ends are always dominant. Whenever the topology has the same amount of free ends with each dipole orientation, like in XSMPs, they tend to connect head-to-tail in pairs, exhausting all the most favourable connections. On the other hand, if the amount of free ends with each dipole orientation is not the same, like in YSMPs, not all of those with a more numerous orientation can find a complementary neighbour. This frustration pushes them to establish less favourable connections. For YSMPs, among the crosslinked particles, the most energetically unfavourable bonds are those between the central junction particle and the two neighbours with dipoles in. Whereas all other pairs of particles are already bonded in a head-to-tail orientation, with little bending of their bonds, the two particles with dipoles in attached to the junction point of the YSMP repeal each other magnetically, tending to bend their bonds and opening a space to a free end of another SMP to approach. So, when a free end seeks to form a connection and finds no other compatible free end, it finds as a more favourable alternative to perturb one of these weak points in the junction.

The interpretation of the role of free ends discussed above also applies nicely to the rest of systems. For instance, for dispersions of LSMPs with $\mu^2=5$ and any sampled length and density, the connectivity maps (not shown) also have a significant probability only for the connection between free ends. Finally, in the case of weak dipole-dipole interactions, the connectivity maps (not shown) display the same qualitative behaviour for all systems, with the only difference of being more blurry due to the increased role of entropy.
\begin{figure*}[!h]
\centering
\includegraphics[width=16cm]{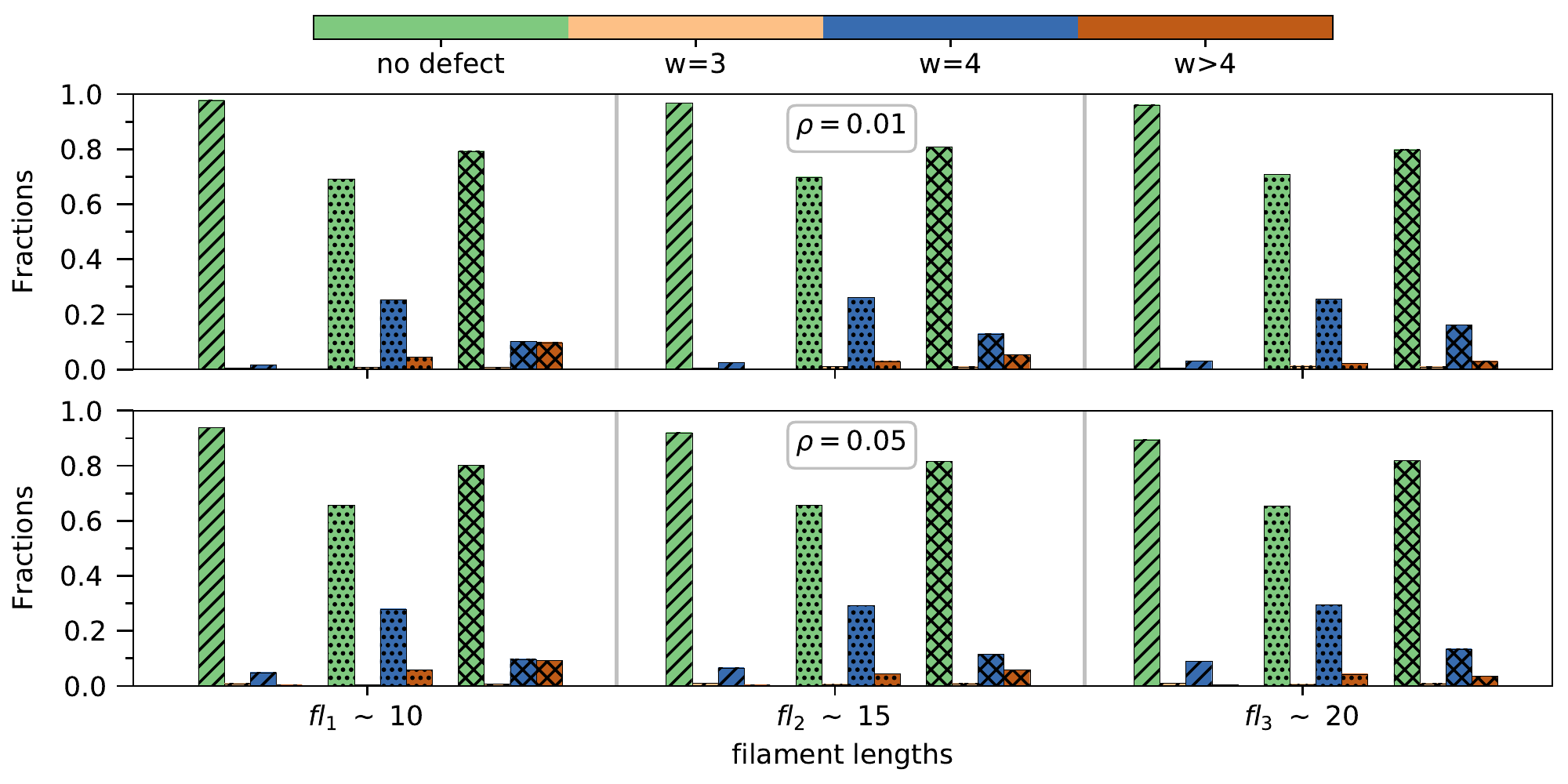}
\caption{\emph{Bar chart $\lambda=5$.} Fractions of particles bonded without the formation of a junction (\emph{green}), with the formation of a $w=3$ defect (\emph{orange}), $w=4$ defect (\emph{blue}) or $w>4$ defect (\emph{red}), for three different values of the filament length. The density value is $\rho=0.01$ for the bar chart on the top, $\rho=0.05$ on the bottom. The hatching of the bars indicates the type of the filaments in the system according to the following identifications: diagonal lines, LSMPs; points, YSMPs; and crosses, XSMPs. Values for simple fluids: for $\rho=0.01$, the fractions are $98.2\%$ particles without junctions, $1.1\%$ with $w=3$, $0.5\%$ with $w=4$ and $0.2\%$ for $w>4$; for $\rho = 0.05$, the fractions are $96\%$ particles without junctions, $2.5\%$ with $w=3$, $0.9\%$ with $w=4$ and $0.6\%$ for $w>4$. In all cases the estimated maximum relative error is lower than 1\%.}
\label{fig:barchartL5}
\end{figure*}

\subsection{Classification} \label{sec-bond}
We use a systematic way of classifying inter-SMPs connections according to their local topology, basing our analysis on the number of neighbours each particle has. The criteria used to define neighbours is based on both energy and distance, as discussed above.


We define \emph{defect} particles as those with more than two bonded neighbours and, if such particles form a close group we address the complete set of defect particles as \emph{defect}. Following the method proposed by Rovigatti \emph{et al.},\cite{2013-rovigatti}, we introduce two parameters to characterise the defects: \emph{s} (the number of defect particles in the defect, namely the size of the defect) and \emph{w} (the number of ways out from the defect).

As previously mentioned, RSMPs rarely form any clusters. So, below we again only report the connection classification for suspensions of LSMPs, YSMPs and XSMPs. We also classify the connections found in the fluids of non-crosslinked dipolar soft spheres.

The bar chart in Figure \ref{fig:barchartL2} (Figure \ref{fig:barchartL5}) presents the fractions of particles involved in defects characterized by $w = 3$, $w=4$ , $w>4$ or $without$ junctions for $\lambda=2$ ($\lambda=5$ respectively).
For each value of particle number density $\rho$ and interaction strength $\lambda$, data are normalised by the total number of particles in the system. 
Moreover, for each value of the filament length, the sum of the fractions of each kind of defect ($w = 3$, $w=4$ , $w>4$ and $without$ junctions) is unity.

For $\lambda =2$ (Fig. \ref{fig:barchartL2}), thermal fluctuations dominate in the system. As a result, firstly, not so many inter-SMPs connections are formed. Secondly, even though such connections have a negative energy, the absolute value of the latter is very small. In this regime, the length of the SMP has the highest impact on the formation of inter-SMP connections. Under this condition, if a connection forms, then with the highest probability it is a $w=4$ junction; fraction of $w=4$ junctions grows with increasing filament length.

In contrast, there are quite a few inter-SMP connections for $\lambda =5$ (Figure \ref{fig:barchartL5}), but basically no defects. From these histograms one can clearly notice that the highest number of defects can be found in YSMPs. It correlates with connectivity probabilities discussed in the previous subsection. Free ends of YSMPs are likely to attach not only to the free ends, but also to the SMP beads in the vicinity of the permanent Y-junction. As an outcome, a X-like configuration can be obtained. We believe that the large amount of such connections is related to the energy gain obtained through X-defect formation. As discussed above, two particles bonded to the central one in a YSMP are in the most energetically unfavourable configuration due to the magnetic repulsion between side-by-side moments, and as such they stretch the crosslinking springs creating a space for another particle to approach. If the free end connects to one of those ``frustrated'' particles, it becomes possible to form two almost head-to-tail pairs.

It is interesting to compare these results with conventional
magnetic fluids containing non-crosslinked nanoparticles. In the studied range of parameters we see basically no junctions. The characteristics for the systems of non-crosslinked dipolar particles can be found in the captions of Figure \ref{fig:barchartL2} for $\lambda = 2$ and Figure \ref{fig:barchartL5} for $\lambda = 5$, respectively. In both cases, more than 95 per cent of connections between the particles in non-crosslinked systems are those without defects.

\begin{table}[!h]
\centering
\small
\begin{tabular}{cccc}
  \toprule
 \emph{Permanent Bonds}   \\
  \midrule
& $fl_1\sim10$ &$fl_2\sim15$&$fl_3\sim20$   \\
  \midrule
  LMP & 0.90&0.93 &0.95 \\
 YMP &   0.90& 0.94  &0.95\\  
  XMP &0.89 & 0.92 &0.95 \\
  \midrule
  Ferrofluids & & 0.0 & \\
 \midrule
 \bottomrule
 \\
\end{tabular}
  \caption{Fractions of permanent bonds per particle for each type of system:  LSMPs, YSMPs, XSMPs, for different values of the SMP lengths. For non-crosslinked systems (ferrofluids) such fraction is zero.}
  \label{permanent-bonds}
\end{table}

\begin{table*}[!h]
\centering
\small
\begin{tabular}{ccccc|ccc}
        \toprule
        \emph{Additional connections}  \\
        \midrule
        &&&\underline{$\lambda=2$}&&&\underline{$\lambda=5$} \\
        && $fl_1\sim10$ &$fl_2\sim15$&$fl_3\sim20$& $fl_1\sim10$ &$fl_2\sim15$&$fl_3\sim20$   \\
        \midrule
        &LMP &0.0038&0.0029&0.0027&0.073&0.048&0.035\\
        $\rho=0.01$&YMP&0.0073&0.0056&0.0047&0.12&0.069&0.057\\
        &XMP&0.023&0.016&0.01&0.20&0.13&0.079\\
        &Ferrofluids&&0.016&&&0.48&\\
        \midrule
        &LMP &0.010&0.0075&0.0066&0.084&0.056&0.043\\
        $\rho=0.05$&YMP&0.015&0.010&0.010&0.13&0.081&0.069\\
        &XMP&0.032&0.022&0.015&0.20&0.14&0.086\\
        &Ferrofluids&&0.073&&&0.49&\\
 \midrule
 \bottomrule
\\
\end{tabular}
  \caption{The fractions of additional connections per particle for each type of system:  LSMPs, YSMPs, XSMPs, for different values of the SMP lengths and systems with non-crosslinked particles. The maximum relative error estimated for these measures is not larger than 4\%.}
  \label{additional-bonds}
\end{table*}
In general, in non-crosslinked systems all particles are potentially available to form connections, whereas in the systems of SMPs, this is by far not the case. In fact, if we look at the data presented in Table \ref{permanent-bonds}, where we collected the fraction of permanent bond per particle in various SMPs, one will see that almost each particle is bonded. In other words, each particle has a bond with the probability of at least 89 per cent independently from the type of SMP.  So, in agreement with connectivity maps (Fig. \ref{fig:connmaps}) discussed above, only few particles in SMPs are participating in the formation of inter-SMP connections. However, for $\lambda =5$, from the analysis of cluster sizes (Fig. \ref{fig:clust-histograms} ) it clearly follows that SMPs do self-assemble, and as shown in Fig. \ref{fig:barchartL5} even form defects. For this reason, we decided to also look at the probability for particles in the SMPs to form a connection. The results are summarised in Table \ref{additional-bonds}. For $\lambda = 2$, it is clearly seen that the probability for a particle to form a connection is rather low, especially for $\rho=0.01$, both for the systems of non-crosslinked particles and for the dispersions of SMPs. For $\rho = 0.05$, the fraction of connections is higher and the largest value is observed for a system without crosslinkers. The second largest fraction of connections per particle is found for XMPs.  This tendency holds also for $\lambda =5$. Even though it might seem that the fraction of connections per particle is much higher for non-crosslinked systems, one should compare this results to those from Table \ref{permanent-bonds}, where the number of permanent bonds is shown. Basically, looking at the fraction of connections per particle in XSMPs systems for $\lambda =5$ and $\rho=0.05$, it is simply striking to find it only three times smaller than the one for ferrofluids: it clearly indicates that SMPs are effectively more fervent to self-assemble than non-crosslinked particles at the same conditions. 

\section{Conclusions} \label{sec-conc}

In this manuscript we employed molecular dynamics computer simulations to describe the self-assembly of magnetic supracolloidal polymers of different topologies: linear, ring-like, Y- and X-shaped. 

The study presented here is two-fold. First of all, through comparing self-assembly of SMPs to that of non-crosslinked magnetic soft spheres, we investigated the influence of crosslinkers on the cluster formation. It turned out that the presence of crosslinkers affects quantitatively the formation of new interparticle bonds, but does not alter the type of them. Namely, for non-crosslinked systems, LSMPs, YSMPs and XSMPs, for the studied range of parameters, the most probable type of the bond is that leading to the linear segment, the second probable connection results in the formation of X-type junction. Interestingly enough, the branching is more pronounced in solutions of SMPs than in ordinary dipolar soft sphere systems. Secondly, we found that the topology of SMPs plays a crucial part in the self-assembly. Thus, RSMPs are not exhibiting any self-assembly in the studied range of parameters. LSMPs do cluster, but the cluster-size distribution resembles strongly that of non-crosslinked dipolar soft spheres, showing and exponential decay of the cluster size. In contrast to the previous two types of SMPs, YSMPs and XSMPs show bimodal cluster-size distributions, with one of the peaks corresponding to single SMPs and the other to very large clusters. The latter aggregates can contain up to 95 - 97 per cent of all SMPs from the small samples we explored. In order to underline the analogy between the molecular valency and the number of SMPs free ends, we additionally analysed the most active participants of the inter-SMP self-assembly. We found that for LSMPs and XSMPs the free ends are the main participants in new connection formation, whereas for YSMPs, also the beads around the junction are able to form an extra connection. The latter can be explained by the weakness of the magnetic interaction of the crosslinked beads forming Y-junction. As a result, attaching a free end to one of them might lead to a creation of a less frustrated dipolar configuration.

Suspensions of SMPs offer a rich variety of self-assembly scenarios depending on the topology and length of the building blocks. The next step, we are currently working on, is to investigate magnetic response of these systems both to weak and strong externally applied fields.

\section{Acknowledgements}
This research has been supported by the Russian Science Foundation Grant No.17-72-10145. P.A.S and S.S.K acknowledge support from the Austrian Research Fund (FWF), START-Projekt Y 627-N27. S.S.K. and M.R. also acknowledge support from the ETN-COLLDENSE (H2020-MSCA-ITN-2014, Grant No. 642774). T.S. acknowledges FIS20015-63628-C2-2-R (AEI/FEDER,UE). We also thank Vladimir Zverev for useful discussion and helping with visualization. Computer simulations were performed at the Ural Federal University computing cluster.


\end{document}